\documentclass[12pt,letterpaper]{article}
\usepackage{amsfonts}
\usepackage{amsmath}
\usepackage{amssymb}
\usepackage[nosort]{cite}
\usepackage{color}
\usepackage{graphicx}
\usepackage{young}
\usepackage[vcentermath]{youngtab}
\usepackage{multirow}
\usepackage{slashed}

\usepackage{lipsum}

\usepackage{hyperref}
\usepackage[left=2.5cm,right=2.5cm,top=2.5cm,bottom=3cm]{geometry}
\numberwithin{equation}{section}

\def\bea#1\eea{\begin{align}#1\end{align}}
\def\bes #1\ees{\begin{split}#1\end{split}}
\newcommand{\be}{\begin{equation}}
\newcommand{\ee}{\end{equation}}

\def\C#1{{(\ref{#1})}}
\def\com#1#2{{ \left[ #1, #2 \right] }}
\def\acom#1#2{{ \left\{ #1, #2 \right\} }}

\newcommand{\wt}{\widetilde}

\newcommand{\ov}{\overline}
\newcommand{\ul}{\underline}

\begin{document}

\begin{titlepage}

\begin{center}

December 1, 2016
\hfill         \phantom{xxx}  EFI-16-19

\vskip 2 cm {\Large\textbf{ Supergravity Backgrounds for Four-Dimensional Maximally Supersymmetric Yang-Mills }}

\vskip 1.25 cm {\bf Travis Maxfield\footnote{email: \tt{maxfield@uchicago.edu}
}}

\vskip 0.2 cm
 $^{1}${\it Enrico Fermi Institute, University of Chicago, Chicago, IL 60637, USA}

\end{center}
\vskip 1 cm

\begin{abstract}
\baselineskip=18pt
In this note, we describe supersymmetric backgrounds for the four-dimensional maximally supersymmetric Yang-Mills theory. As an extension of the method of Festuccia and Seiberg to sixteen supercharges in four dimensions, we utilize the coupling of the gauge theory to maximally extended conformal supergravity. Included among the fields of the conformal supergravity multiplet is the complexified coupling parameter of the gauge theory; therefore, backgrounds with spacetime varying coupling---such as appear in F-theory and Janus configurations---are naturally included in this formalism. We demonstrate this with a few examples from past literature. 
\end{abstract}

\end{titlepage}

\tableofcontents

\section{Introduction}
Given a quantum field theory, an essential and often fruitful ingredient in its study is provided by its couplings to background fields. In the context of supersymmetric theories, the question of which backgrounds preserve some supersymmetry has lately received much attention, especially in conjunction with localization\cite{Pestun:2007rz,Pestun:2016jze}. A powerful, general technique for addressing this question was put forward in\cite{Festuccia:2011ws}, which relied on coupling the given field theory to an off-shell formulation of supergravity. Upon taking a limit in which the supergravity fields become non-dynamical, these fields---which include the metric as well as other fields as needed to complete the supermultiplet---give background couplings. This technique has been applied in a number of cases in different dimensions and with different amounts of supersymmetry. For a recent review, see \cite{Dumitrescu:2016ltq}.

In this paper we extend the application of the technique in\cite{Festuccia:2011ws} to sixteen supercharges in four-dimensions by coupling the $\mathcal{N} = 4$ Super Yang-Mills (SYM) theory to off-shell conformal supergravity\cite{Bergshoeff:1980is,deRoo:1984zyh,deRoo:1985np}. A qualitatively new feature that arises in this case is the inclusion of the complexified coupling paramter, $\tau$, of $\mathcal{N} = 4$ SYM among the fields of the conformal supergravity multiplet\cite{Howe:1981qj}. Therefore, in this case, supersymmetric backgrounds with spacetime-varying $\tau$ are treated on the same footing as metrics and $R$-symmetry gauge fields. 

Backgrounds with spacetime-varying $\tau$ have arisen in the physics of D3-branes probing non-trivial F-theory vacua, where the field theory $\tau$ descends from the Type IIB axio-dilaton, and the constraints imposed by supersymmetry descend also from Type IIB supergravity and the kappa symmetry of the brane probe\cite{Harvey:2007ab,Martucci:2014ema}. Varying $\tau$ also appears in Janus configurations\cite{D'Hoker:2006uv,Gaiotto:2008sd,Kim:2008dj,Chen:2008tu,Kim:2009wv} in which $\tau$ varies along a single spacetime direction and interpolates between two asymptotic values. In both cases, supersymmetry constrains the spacetime dependence of $\tau$ and further requires the field theory to be modified by additional couplings. Furthermore, in both cases $\tau$ can possess localized discontinuities, or defects, whose worldvolumes support interesting field theories. These field theories can be chiral\cite{Harvey:2007ab,Martucci:2014ema}, possess interesting S-duality transformations\cite{Gaiotto:2008sa}, and be relevant to the study of the AGT relation\cite{Hosomichi:2010vh}.

Both of these examples---in addition to some novel Janus configurations which we will describe---are included among the backgrounds found by coupling to conformal supergravity, as we will demonstrate in a few cases. While we have not attempted a systematic classification of supersymmetric backgrounds, we expect this method to yield a broad generalization of the known backgrounds with varying $\tau$. It is our hope that such backgrounds---combined with techniques like localization---may shed new light on these and the associated defect field theories.

Before proceeding, we would like to comment on a possible relation between the method of\cite{Festuccia:2011ws} and the physics of branes probing flux vacua. This has already been alluded to above, and it has been suggested previously in\cite{Maxfield:2015evr,Triendl:2015lka}. To make the relation clear, we recall in analogy the geometrization of Witten's topological twist\cite{Witten:1988ze}. Topological twisting can be thought of as a special case of\cite{Festuccia:2011ws}, in which only the metric and a background $R$-symmetry gauge field are non-trivial. For certain field theories, this construction has a geometrical origin as the low energy description of branes wrapping supersymmetric cycles in manifolds of special holonomy\cite{Bershadsky:1995qy,Becker:1995kb}.

As mentioned, the topological twist can be generalized to include the auxiliary fields residing in the off-shell supermultiplets. Likewise, the brane construction can be generalized to include to supergravity flux degrees of freedom. One is naturally lead to wonder what relation these generalizations have to one another, and there is some evidence they are the same, at least in special circumstances. For example, in \cite{Maxfield:2015evr} a comparison was made between the flux couplings on an M5-brane and the auxiliary fields of the off-shell $(2,0)$ conformal supergravity multiplet\cite{Bergshoeff:1999db,Cordova:2013bea}, which led to a proposed mapping between them. 

If true, such a relation should provide new tools to the study of both supersymmetric field theories in non-trivial backgrounds and string theory flux vacua. As one such possibility, the localization calculations of, e.g.,\ \cite{Pestun:2007rz} may be relevant to brane instantons in flux backgrounds. Alternatively, string theory dualities could give new interpretations to the field theory calculations. In this paper, we will not attempt a general proof of the proposed relation between off-shell supergravity and higher-dimensional supergravity with fluxes. However, the supersymmetry conditions we write down can be compared with the couplings of background fluxes to D3-branes in\cite{Marolf:2004jb,Martucci:2005rb}.

This paper is organized as follows. Section~\ref{review} contains a review of the fields and symmetries of $\mathcal{N} = 4$ conformal supergravity. Section~\ref{matter} describes the coupling of these fields to the $\mathcal{N} = 4$ matter multiplet. Lastly, section~\ref{examples} describes some example supersymmetric background configurations, including the cases mentioned previously. Two appendices include, respectively, details about our notation and a dictionary of comparison between our formulae and the original literature on extended conformal supergravity\cite{Bergshoeff:1980is}. 

\section{Review of ${\cal N} = 4$ Conformal Supergravity}\label{review}
Like other conformal supergravity theories,\footnote{For a modern review see~\cite{Freedman:2012zz}.} $\mathcal{N} = 4$ conformal supergravity is a modified version of the gauge theory of its conformal group, $PSU(4|4)$~\cite{Bergshoeff:1980is}. The necessary modifications include the addition of a set of constraints relating together some of the $PSU(4|4)$ gauge fields as well as a number of auxiliary fields to complete the conformal supergravity or ``Weyl" multiplet. These modifications ensure that the algebra of $PSU(4|4)$ transformations on the fields of the theory closes even off-shell, which is necessary for our purposes.

This section will be split into three parts. First, we describe the gauge fields and constraints of the $PSU(4|4)$ gauge theory. Second, we list the auxiliary fields needed to complete the conformal supergravity multiplet, with an emphasis on the moduli of the theory. Lastly, we will present the supersymmetry variations of the fermions of the theory, here interpreted as sufficient conditions for a supersymmetric background configuration of the $\mathcal{N} = 4$ matter theory. This latter part will include a description of the symmetries and transformations of the independent fields of the theory. Apart from some aesthetic modifications, the presentation of this section is an abridging of portions of~\cite{Bergshoeff:1980is}.

It is worthwhile pausing here to make some comments on our conventions. More detailed exposition can be found in Appendix~\ref{app:conventions}. We will primarily work in Lorentzian signature, for which the supersymmetries transform in the $(\mathbf{2}, \mathbf{4}) \oplus (\ov{\mathbf{2}}, \ov{\mathbf{4}})$ representation of $Spin(1,3) \times SU(4)_R$. We denote orthonormal frame indices with lowercase Latin letters from the beginning of the alphabet ($a, b, \ldots$), while spacetime indices are lowercase Greek letters from the middle of the alphabet ($\mu, \nu, \ldots$). We will almost always suppress spacetime spinor indices, though when necessary they are denoted with lowercase Greek letters from the beginning of the alphabet ($\alpha, \beta, \ldots$). Lastly, lowercase Latin indices from the middle of the alphabet ($i,j, \ldots$) refer to the four-dimensional representations of $SU(4)_R$ depending on whether they are superscripts or subscripts.

We denote the conjugate of any complex field with an overline. For a spinor field $\psi$, conjugation carries some extra baggage, i.e.\ $\ov \psi = B \psi^*$, where $B$ is a particular matrix discussed further in Appendix~\ref{app:conventions}. When $\psi$ is a four-dimensional Weyl spinor, $\ov \psi$ transforms as a spinor of the opposite chirality as well as in the conjugate representation of any internal symmetry, such as $SU(4)_R$. When listing the independent fields of the Weyl multiplet, we won't list both a field and its conjugate, instead only pointing out those fields that are complex.

\subsection{Gauge fields and constraints}

\begin{table}[h!]
\centering
\begin{tabular}{|c|c|c|c|c|}
\hline
Symmetry & Generators & Gauge Field & Type & Restriction \\
\hline
\multirow{2}{*}{Translations} & \multirow{2}{*}{$P_a$} & \multirow{2}{*}{$e_\mu^{a}$}& \multirow{2}{*}{boson} & \multirow{2}{*}{coframe} \\
& & & & \\
\hline
\multirow{2}{*}{Lorentz} & \multirow{2}{*}{$M_{ab}$} & \multirow{2}{*}{$\omega_{\mu}^{ab}$} & \multirow{2}{*}{boson} &\multirow{2}{*}{ spin-connection} \\
& & & & \\
\hline
$SU(4)$& \multirow{2}{*}{$U^{i}{}_j$} & \multirow{2}{*}{$V^{i}_{\mu\phantom{i} j}$} & \multirow{2}{*}{boson} & $\left(V^i_{\mu \phantom{j} j}\right)^\ast = -V^j_{\mu \phantom{j} i}$\\
R-symmetry & & & & $V^{i}_{\mu \phantom{i}i}= 0 ~$ \\
\hline
\multirow{2}{*}{Dilation} &\multirow{2}{*}{ $D$} & \multirow{2}{*}{$b_{\mu}$} & \multirow{2}{*}{boson} & \multirow{2}{*}{--}\\
 & & & & \\
\hline
\multirow{2}{*}{Special Conformal} &  \multirow{2}{*}{$K_{a}$} &  \multirow{2}{*}{$f^{a}_{\mu}$} &  \multirow{2}{*}{boson} &  \multirow{2}{*}{--} \\
 & & & & \\
\hline
\multirow{2}{*}{Supersymmetry} & \multirow{2}{*}{$Q^{ i}$ }& \multirow{2}{*}{$ \psi^i_{{ \mu}}$} & \multirow{2}{*}{fermion} & $\gamma_{(5)} \psi_\mu^i = \psi_\mu^i \phantom{-}$\\
& & & & complex \\
\hline
Conformal & \multirow{2}{*}{$S_{i}$ }& \multirow{2}{*}{$ \phi_{{\mu i}}$} & \multirow{2}{*}{fermion} &$\gamma_{(5)} \phi_{\mu i} = \phi_{\mu i}\phantom{-}$\\
Supersymmetry& & & & complex\\
\hline
\end{tabular}
\caption{Gauge fields of four-dimensional $\mathcal{N} = 4$ conformal supergravity.}
\label{tab:gaugefields}
\end{table}

The set of symmetry generators and the corresponding gauge fields of the $PSU(4|4)$ gauge theory are shown in Table~\ref{tab:gaugefields}. Not all of these are independent, however, due to the presence of constraints. In a bosonic background, these constraints imply:
\be
\bes
\omega_\mu^{ab} &= \hat \omega_\mu^{ab} + 2 e_\mu^{[a} b^{b]}, \cr
f_\mu^\mu &= -{1\over 12} R(\omega),
\ees
\ee
where $\hat \omega_\mu^{ab}$ is the torsion-free spin-connection associated to the coframe $e_\mu^a$, and $R(\omega)$ is the Ricci scalar of the modified spin-connection $\omega_\mu^{ab}$. Additionally, while we have no need of the precise constraints, the field $\phi_{\mu i}$ and its conjugate are not independent; they are related to $\psi^i_\mu$ and its conjugate.

It will be convenient for us to fix the special conformal gauge symmetry by making the choice $b_\mu = 0$. Under a combined $D$ and $K^a$ transformation, with paramaters $\Lambda_D$ and $\Lambda_K^a$, $b_\mu$ transforms as
\be
\delta b_\mu = \partial_\mu \Lambda_D + \Lambda_K^a e_{\mu a}.
\ee
So, in order to maintain this gauge choice we must fix
\be
\Lambda_K^a = - e^{a \mu} \partial_\mu \Lambda_D.
\ee
The only fields of the Weyl multiplet, including the auxiliary fields introduced in the next subsection, that are charged under special conformal transformations are $b_\mu$, $\omega_\mu^{ab}$ and $f_\mu^a$. Therefore, in this gauge, the latter two fields have modified Weyl transformations, with
\be
\delta \omega_\mu^{ab} = -2 e_\mu^{[a}e^{b] \nu} \partial_\nu \Lambda_D.
\ee
The modified transformation of $f_\mu^\mu$ can be deduced from the constraint equation. Note also that in this gauge $\omega_\mu^{ab} = \hat{ \omega}_\mu^{ab}$, the spin-connection associated to $e_\mu^a$. We will continue to refer to this as $\omega_\mu^{ab}$.

\subsection{Auxiliary fields} 
For the algebra of $PSU(4|4)$ transformations to close on-shell requires the inclusion of a set of auxiliary fields in addition to the gauge fields in Table~\ref{tab:gaugefields}. The presence of these fields actually enhances the symmetry of the theory beyond that of $PSU(4|4)$, including as well a global $SL(2,\mathbb{R})$ and a local, chiral $U(1)$ symmetry~\cite{Bergshoeff:1980is}. In Table~\ref{tab:allfields}, we list the independent fields of this theory, as well as their representations and charges under the $SU(4)_R$, $U(1)$, and Weyl symmetries.

\begin{table}[h]
\centering
\begin{tabular}{|c|c|c|c|c|c|c|}
\hline
Field & Type & Properties & $SU(4)_R $ & Weyl Weight & $U(1)$ & \\
\hline
$\tau$ & boson & $\mathrm{Im}~\! \tau > 0$ & $\mathbf{1}$ &  $0$ & $0$ & $\ast$ \\
\hline

$E_{ij}$ & boson & $E_{ij} = E_{ji}$ & $\ov{\mathbf{10}}$ & $1$ & $-1$ & $\ast$ \\
\hline
\multirow{2}{*}{$T_{ab}{}^{ij}$} & \multirow{2}{*}{boson} & $T_{ab}{}^{ij} = -T_{ba}{}^{ij} = -T_{ab}{}^{ji}$ & \multirow{2}{*}{$\mathbf{6}$} & \multirow{2}{*}{$1$} & \multirow{2}{*}{$-1$} & \multirow{2}{*}{$\ast$}\\
& & $T_{ab}{}^{ij}={i \over 2} \epsilon_{ab}{}^{cd} T_{cd}{}^{ij}\qquad~~\! $ & & && \\
\hline
\multirow{3}{*}{$D^{ij}{}_{kl}$} & \multirow{3}{*}{boson} & $D^{ij}{}_{kl} = -D^{ji}{}_{kl} = -D^{ij}{}_{lk} \qquad~\!$ & \multirow{3}{*}{$\mathbf{20}$} & \multirow{3}{*}{$2$} & \multirow{3}{*}{$0$} & \\
& & $\left(D^{kl}{}_{ij}\right)^\ast = D^{ij}{}_{kl} = {1 \over 4} \epsilon^{ijmn}\epsilon_{klpq}D^{pq}{}_{mn}$ & & & & \\
& & $D^{ij}{}_{kj} = 0\phantom{\epsilon^{ijmn}\epsilon_{klpq}D^{pq}{}_{mn}} \qquad~~\!$ & & & &\\
\hline
\multirow{2}{*}{$V^{i}_{\mu\phantom{i} j}$} & \multirow{2}{*}{ boson} & $\left(V^i_{\mu \phantom{j} j}\right)^\ast = -V^j_{\mu \phantom{j} i}$ & \multirow{2}{*}{$\mathbf{15}$} & \multirow{2}{*}{$0$} & \multirow{2}{*}{$0$} & \\
& & $V^{i}_{\mu\phantom{i} i} = 0 \phantom{V^{j}_{\mu\phantom{j} i}}$ & & & & \\
\hline
$e^a_\mu$ & boson & coframe & $\mathbf{1}$ & $-1$ & $0$ & \\
\hline
$\Lambda_{ i}$ & fermion & - & $\ov{\mathbf{4}}$ &  ${1 \over 2}$ & $-{3 \over 2}$ & $\ast$ \\
\hline
\multirow{2}{*}{$\chi^{ij}{}_k$} & \multirow{2}{*}{fermion} & $\chi^{ij}{}_k = - \chi^{ji}{}_k$& \multirow{2}{*}{$\ov{\mathbf{20'}} $}& \multirow{2}{*}{${3 \over 2}$} & \multirow{2}{*}{$-{1 \over 2}$} &\multirow{2}{*}{$\ast$} \\
& & $\chi^{ij}{}_i = 0 \qquad \!$ & & & & \\
\hline
$\psi^i_{ \mu}$ & fermion& gravitino & $\mathbf{4}$ & $-{1 \over 2}$ & $-{1 \over 2}$ & $\ast$ \\
\hline

\end{tabular}
\caption{Independent matter and gauge fields of four-dimensional $\mathcal{N} = 4$ conformal supergravity. Fields marked with a $\ast$ on the right are complex.}
\label{tab:allfields}
\end{table}

Only one field transforms under the $SL(2,\mathbb{R})$ symmetry, and that is a scalar doublet called $\Phi_{\ul \alpha}$, $\ul \alpha = 1, 2$ in~\cite{Bergshoeff:1980is}. This field parameterizes the coset $SL(2,\mathbb{R})/U(1)$, which is isomorphic to the upper half-plane. This is also the space of complexified couplings, $\tau$, of the $\mathcal{N} = 4$ matter theory. Indeed, after a convenient $U(1)$ gauge-fixing, described in Appendix~\ref{app:gaugefixing}, we can identify $\Phi_{\ul \alpha}$ with $\tau$.

The residual symmetry after this gauge-fixing includes the $PSL(2,\mathbb{R})$ transformations of $\tau$:
\be
\tau \to {a \tau + b \over c\tau + d}, \quad a, b, c, d, \in \mathbb{R}, \quad ad - bc =1.
\ee
However, to maintain the gauge choice, a compensating $U(1)$ transformation is required, which acts on charge $q$ fields by multiplication with
\be
e^{i q \theta} = \left({|c \tau + d| \over c\tau +d }\right)^q.
\ee
Owing to the fermions in the theory having half-integral $U(1)$ charges $q$, we find that the residual symmetry is not just $PSL(2,\mathbb{R})$ but an extension by the the operation $(-1)^F$, where $F$ is the fermion number operator. This is the same conclusion reached in~\cite{Kapustin:2006pk,Harvey:2007ab} regarding the quantum duality group. We will present the transformations of the conformal fields as well as their couplings to the Yang-Mills fields only in gauge-fixed form.

\subsection{Supersymmetry conditions}
The entire superconformal algebra realized on the fields in Table~\ref{tab:allfields} can be found in~\cite{Bergshoeff:1980is}. We will only need the $Q$- and $S$-supersymmetry transformations, with respective parameters $\epsilon$ and $\eta$, of the fermionic fields. 
\be\label{eq:susyconditions}
\bes
\delta \Lambda_i &= i {\partial_\mu \ov \tau \over \mathrm{Im}~\! \tau } \gamma^\mu \ov \epsilon_i + E_{ij}\epsilon^j + T_{abij}\gamma^{ab} \epsilon^j, \cr
\delta \chi^{ij}{}_k &= -{1 \over 2} \mathcal{D}_\mu T_{ab}{}^{ij} \gamma^{ab} \gamma^\mu \ov\epsilon_k - R_{ab}{}^{[i}{}_k \gamma^{|ab|} \epsilon^{j]}- {1 \over 2} \epsilon^{ijlm} \mathcal{D}_\mu E_{k l} \gamma^\mu \ov \epsilon_m + D^{ij}{}_{k l} \epsilon^l +{1 \over 6} T_{ab kl} \ov E^{l[i} \gamma^{|ab|} \epsilon^{j]}  \cr
&\phantom{=}~     + { 1\over 2} E_{kl}\ov E^{l[i} \epsilon^{j]} - i {\partial_\mu \ov \tau \over \mathrm{Im}~\! \tau} \ov T_{ab}{}^{ij}\gamma^\mu \gamma^{ab} \ov\epsilon_k +{1 \over 2} T_{ab}{}^{ij} \gamma^{ab} \eta_j  -{ 1\over 2} \epsilon^{ijlm} E_{kl}\eta_m \cr
&\phantom{=}~ - (\textrm{traces}), \cr
\delta \psi^i_\mu & =  2 \mathcal{D}_\mu \epsilon^i - {1 \over 2} T_{ab}{}^{ij} \gamma^{ab} \gamma_\mu\ov  \epsilon_j - \gamma_\mu \ov\eta^i.
\ees
\ee
The notation ``$-(\mathrm{traces})$" refers to removing any terms proportional to the $SU(4)_R$ invariant tensor $\delta^i{}_j$. The first variation, that of $\Lambda_i$, carries some resemblance to the dilatino variation in Type IIB supergravity, and we will refer to it with that name. In fact, from the perspective espoused in the introduction, in which the conformal supergravity fields may be identified with Type IIB supergravity fields and fluxes, it is natural to think this constraint is actually descended from the dilatino variation. However, the precise relation remains to be explored. We will refer to the second and third variations as ``auxiliary" and ``gravitino", respectively. The covariant derivatives and curvatures appearing above are, explicitly,
\be
\bes
\mathcal{D}_\mu \epsilon^i &= \partial_\mu \epsilon^i   -{1 \over 4}\omega_\mu{}^{ab} \gamma_{ab} \epsilon^i - V_{\mu\phantom{i} j}^i \epsilon^j  -{1 \over 2} a_\mu(\tau) \epsilon^i , \cr
\mathcal{D}_\mu T_{ab}{}^{ij} &= \partial_\mu T_{ab}{}^{ij}  + 2 \omega_{\mu \phantom{c} [a}^{\phantom{\mu} c}T_{b]c}{}^{ij} + 2 V_{\mu \phantom{i} k}^{[i} T_{ab}{}^{j] k} - a_\mu(\tau) T_{ab}{}^{ij}, \cr
\mathcal{D}_\mu E_{ij} &= \partial_\mu E_{ij}  + 2 V_{\mu \phantom{k} (i}^k E_{j) k} - a_\mu(\tau) E_{ij}, \cr
R_{ab}{}^i{}_j &= 2\partial_{[\mu} V_{\nu] \phantom{i} j}^i - 2V_{[\mu \phantom{i} |k|}^i V_{\nu] \phantom{k} j}^k, \cr
a_{\mu}(\tau) &= -i{ \partial_\mu \left( \tau + \ov \tau \right) \over 4 \mathrm{Im}~\! \tau}.
\ees
\ee

\section{Coupling to Matter}\label{matter}
There is only one matter multiplet with $\mathcal{N} = 4$ supersymmetry, which is the vector multiplet. The component fields of this multiplet are shown in Table~\ref{tab:4dYMfields}. Each of the component fields is valued in the Lie algebra of a compact, internal symmetry group $G$, which we choose to be simple.

\begin{table}[h]
\centering
\begin{tabular}{|c|c|c|c|c|c|}
\hline
Field & Type & Restrictions & $SU(4)_R $ & Weyl Weight & $U(1)$ \\
\hline
$A_\mu$ & boson & gauge field & $\mathbf{1}$ & $0$  & $0$ \\
\hline
$\psi^i$ & fermion & $\gamma_{(5)} \psi^i = \psi^i$ & $\mathbf{4}$ & ${3 \over 2}$ & ${1 \over 2}$ \\
\hline
$\ov\psi_i$ & fermion & $\gamma_{(5)} \ov\psi_i = -\ov\psi_i$ & $\mathbf{\ov 4}$ & ${3 \over 2}$ & $-{1 \over 2}$ \\
\hline
$\phi^{ij}$ & boson & $\left(\phi^{ij}\right)^* = \phi_{ij} =  {1 \over 2} \epsilon_{ijkl} \phi^{kl}$ & $\mathbf{6} $ & $1$ & $0$ \\
\hline
\end{tabular}
\caption{Fields of an $\mathcal{N} = 4$ matter multiplet.}
\label{tab:4dYMfields}
\end{table}

We furthermore choose conventions for which the generators of $G$ are antihermitian. Then, infinitesimal gauge transformations with parameter $\Lambda$ act as
\be
\bes
A_\mu &\to A_\mu -\partial_\mu \Lambda - \com{A_\mu}{\Lambda}, \cr
\psi^i &\to \psi^i + \com{\Lambda}{\psi^i}, \cr
\phi^{ij} &\to \phi^{ij} + \com{\Lambda}{\phi^{ij}},
\ees
\ee
We normalize the negative definite bilinear form, $``\mathrm{Tr}"$, on the Lie algebra of $G$ to
\be
\mathrm{Tr}~ T^A T^B = -\delta^{AB}, \quad T^A \in \mathrm{Lie}(G).
\ee

The coupling of the conformal supergravity multiplet to the matter fields of $\mathcal{N} = 4$ Yang-Mills was worked out in~\cite{deRoo:1984zyh,deRoo:1985np,Bergshoeff:1985ms}. The transformation of the matter fields under conformal supegravity is determined by requiring that the algebra of such transformations closes when the matter fields are taken on-shell. This also serves to determine the equations of motion, and from this an action can be derived. Before presenting the results of~\cite{deRoo:1984zyh,deRoo:1985np,Bergshoeff:1985ms}, we will settle on a parameterization of the real and imaginary parts of $\tau$:
\be
\tau = {\theta \over 2\pi} + {4\pi i \over g_{YM}^2}.
\ee 

The action of supersymmetry on the matter fields is given by
\be
\bes
\delta A_\mu &= g_{YM} \left( \epsilon^i \gamma_\mu \ov\psi_i + \ov\epsilon_i \gamma_\mu \psi^i \right) , \cr
\delta \psi^i & = -{1\over 2g_{YM} }F_{\mu \nu} \gamma^{\mu \nu} \epsilon^i - 2 \mathcal{D}_\mu \phi^{ij} \gamma^\mu \ov\epsilon_j + \ov E^{ij}\phi_{jk} \epsilon^k - {1\over 2} T_{ab}{}^{jk}\phi_{jk} \gamma^{ab} \epsilon^i  - 2g_{YM}  \com{\phi^{ij}}{\phi_{jk}} \epsilon^k - 2 \phi^{ij} \eta_j, \cr
\delta \phi^{ij} &= 2\epsilon^{[i} \psi^{j]} - \epsilon^{ijkl} \ov\epsilon_k \ov\psi_l.
\ees
\ee
The field strength and covariant derivatives are
\be
\bes
F_{\mu \nu} &= \partial_\mu A_\nu - \partial_\nu A_\mu + \com{A_\mu}{A_\nu}, \cr
\mathcal{D}_\mu \phi^{ij} &= \partial_\mu \phi^{ij} + 2V_\mu^{[i}{}_k \phi^{j]k} + \com{A_\mu}{\phi^{ij}},\cr
\mathcal{D}_\mu \psi^i & = \partial_\mu \psi^i - {1 \over 4}\omega_\mu^{ab} \gamma_{ab} \psi^i - V_\mu^{i}{}_j \psi^j + {1 \over 2} a_\mu(\tau)\psi^i  + \com{A_\mu}{\psi^i}.
\ees
\ee
The action which is invariant under these supersymmetry transformations is
\be
\bes
S_{A} &=\int \!\! d^4x \sqrt{|g|} ~\mathrm{Tr}~ \Big( { i \theta \over 32\pi^2} F_{\mu \nu}\left( \ast F\right)^{\mu \nu} -{1\over 4g_{YM}^2 } F_{\mu \nu}F^{\mu \nu} \Big), \cr
S_{\phi} &=- \int \!\!d^4 x\sqrt{|g|} ~ \mathrm{Tr}~ \Big(~ {1 \over 2} \mathcal{D}_\mu \phi^{ij} \mathcal{D}^\mu \phi_{ij} - {1 \over 2} \phi_{ij} \left( M_\phi \right)^{ij}{}_{kl} \phi^{kl}  +{1 \over g_{YM}} F^{\mu \nu }\left( T_{\mu \nu}{}^{ij} + \ov T_{\mu \nu}{}^{ ij }\right) \phi_{ij} \Big), \cr 
S_{\psi} & = - \int \!\! d^4x\sqrt{|g|} ~\mathrm{Tr}~ \Big(\ov \psi_i \gamma^\mu \mathcal D_\mu \psi^i - {1 \over 2}  \psi^i \left( M_\psi \right)_{ij} \psi^j  - {1 \over 2}  \ov \psi_i \left( \ov M_{\ov \psi} \right)^{ij} \ov \psi_j \Big), \cr
S_{int} &= -\int \!\! d^4x \sqrt{|g|}~\mathrm{Tr}~ \Big( g_{YM} \phi_{ij} \com{\psi^i}{\psi^j} + g_{YM} \phi^{ij} \com{\ov\psi_i}{\ov\psi_j} - {g_{YM}^2 \over 2} \com{\phi^{ij}}{\phi_{jk}}\com{\phi^{kl}}{\phi_{li}} \cr
&\phantom{=-\int \!\! d^4 x~ \sqrt{|g|}\mathrm{Tr}~ \Big(}~ -{g_{YM} \over 3}\ov E^{kl} \phi^{ij} \com{\phi_{ik}}{\phi_{jl}}  - {g_{YM} \over 3} E_{kl} \phi_{ij} \com{\phi^{ik}}{\phi^{jl}} \Big), \cr
S &= S_A + S_\phi + S_\psi + S_{int}.
\ees
\ee
To simplify the presentation, we have defined the following mass matrices for the bosons and fermions:
\be
\bes
\left( M_\phi \right)^{ij}{}_{kl} &= - T_{ab}{}^{ij} T^{ab}{}_{ kl} - \ov T_{ab}{}^{ij}\ov T^{ab}{}_{ kl}  + {1 \over 2}D^{ij}{}_{kl}
-{1 \over 12} \delta^{[i}{}_k \delta^{j]}{}_l \left( \ov E^{mn} E_{mn} -  {\partial_\mu \tau \partial^\mu \ov \tau \over \left(\mathrm{Im}~\! \tau\right)^2} + R(\omega) \right), \cr
\left(M_\psi \right)_{ij} &= {1 \over 2} E_{ij} - {1 \over 2} T_{ab ij} \gamma^{ab}, \cr
\left(\ov M_{\ov\psi} \right)^{ij} &= {1 \over 2} \ov E^{ij} - {1 \over 2}  \ov T_{ab}{}^{ij}\gamma^{ab}.
\ees
\ee

\section{Example Solutions}\label{examples}
We will not attempt to describe the most general solutions to~\C{eq:susyconditions}; instead, we will focus on casting a few known examples of supersymmetric configurations into the language of background conformal supergravity. We choose to focus on two separate classes of configurations with a varying $\tau$ parameter. The first, relevant to D3-branes in F-theory, can be found in\cite{Harvey:2007ab,Martucci:2014ema}; the second are known as Janus configurations\cite{D'Hoker:2006uv,Gaiotto:2008sd,Kim:2008dj,Chen:2008tu,Kim:2009wv}.

\subsection{F-theory examples}
The configurations considered in\cite{Harvey:2007ab,Martucci:2014ema} are characterized by the four-dimensional spacetime being K\"ahler and the $\tau$ parameter being holomorphic in complex coordinates $z_\alpha$, $\alpha = 1, 2$. Holomorphy of $\tau$ is sufficient to solve the dilatino condition without the need for non-zero $E_{ij}$ and $T_{ab}{}^{ij}$. Chirality of the supersymmetry parameters $\epsilon^i$ and $\ov \epsilon_i$ implies
\be
\gamma^{z_\alpha} \epsilon^i = 0, \quad  \gamma^{\ov z_\alpha} \ov \epsilon_i  = 0.
\ee
Therefore, the $\tau$-dependent part of the dilatino constraint,
\be
\delta \Lambda_i = i {\partial_\mu \ov \tau \over \mathrm{Im}~\! \tau } \gamma^\mu \ov \epsilon_i + \ldots,
\ee
vanishes for holomorphic $\tau$ (or, equivalently, anti-holomorphic $\ov \tau$).

The gravitino condition, absent the $T_{ab}{}^{ij}$ field, is the standard topological twist with the addition of a $U(1)$ gauge field---$a_\mu$ in~\C{eq:susyconditions}---which vanishes for constant $\tau$. Lastly, the auxiliary condition relates $D^{ij}{}_{kl}$, which couples to the field theory as a bosonic mass term, to the curvature of the $R$-symmetry gauge field. Such a mass term appears in the standard topological twist, e.g.\ \cite{Vafa:1994tf,Harvey:1999as}, and can be seen to arise from the bulk Killing spinor condition in 10- or 11-dimensional supergravity\cite{Maxfield:xxxx}.

\subsection{Janus examples}
To compare with\cite{D'Hoker:2006uv,Gaiotto:2008sd,Kim:2008dj,Chen:2008tu,Kim:2009wv}, we will look for solutions to~\C{eq:susyconditions} satisfying the following input assumptions: first, the solutions exist in Minkowski spacetime with the trivial metric, and they preserve a three-dimensional Poincare symmetry. Therefore, the coupling $\tau$ is allowed to vary only along a one-dimensional subspace parameterized by $y$, i.e.\ $\tau = \tau(y)$. We choose coordinates to align this direction with that of $x^3$, so the metric is 
\be
ds^2 = \eta_{\mu \nu} dx^\mu dx^\nu = \eta_{\ul {\mu \nu}} dx^{\ul \mu}dx^{ \ul \nu} + dy^2,
\ee
where $x^{\ul \mu}$, $\ul \mu = 0, 1, 2$ are three-dimensional coordinates and $\eta_{\ul {\mu \nu}}$ is the three-dimensional Minkowski metric. Preservation of the three-dimensional Poincare group disallows a non-trivial profile for the background field $T_{ab}{}^{ij}$. Thus, we seek solutions with (possibly) non-trivial profiles for the following fields in addition to $\tau$:
\be
E_{ij}(y), \quad V^i_y{}_j(y), \quad D^{ij}{}_{kl}(y).
\ee

We can simplify our starting point further by picking an $SU(4)_R$ gauge in which $V_y^i{}_j = 0$. More general background configurations can be achieved by applying a local $SU(4)_R$ rotation to the remaining fields after a solution is found.

The gravitino variation can be split:
\be
\bes
\partial_{\ul\mu} \epsilon^i  &= 0, \cr
\partial_y \epsilon^i  + i {g^2 \theta'(y) \over 32\pi^2} \epsilon^i &= 0,
\ees
\ee
where we assume $\ov\eta^i = 0.$\footnote{Non-zero $\ov\eta$ correspond to conformal supersymmetries. We will not discuss this topic, because it doesn't differ substantially from the presentation in\cite{D'Hoker:2006uv,Gaiotto:2008sd}.} In this case, we can straightforwardly solve for $\epsilon^i$:
\be
\epsilon^i(y) = e^{-{i\over 2} \psi(y)} \xi^i, \quad \psi'(y) = {g_{YM}^2 \theta'(y) \over 16\pi^2} = i a_y(\tau),
\ee
where $\xi^i$ is a constant, Weyl spinor. This does not yet specify which, if any, spinors yield supersymmetries; instead, this condition only says that if a supersymmetry exists, it must be generated by a spinor of the above form. The dilatino condition, which we come to next, will constrain the allowed supersymmetries.

In terms of the constant spinors $\xi^i$, the dilatino condition is
\be
e^{i\psi}\left( {\ov \tau' \over \mathrm{Im} ~\tau} \right) \gamma_3 \ov \xi_i = iE_{ij}\xi^j.
\ee
Define
\be
iE_{ij} = e^{i\psi}\left( {i \ov \tau' \over \mathrm{Im} ~\tau} \right) e_{ij}.
\ee
In terms of which, we have the equation
\be
e^{i\psi}\left( {i \ov \tau' \over \mathrm{Im} ~\tau} \right) \left( \gamma_3 \ov \xi_i - e_{ij} \xi^j \right) = 0.
\ee
Our starting point assumed that $\tau'$ is not everywhere zero, so we must have
\be
\gamma_3 \ov \xi_i = e_{ij} \xi^j,
\ee
where $e_{ij}$ is a constant, complex, symmetric matrix.

There is a basis of gamma matrices, detailed in Appendix~\ref{app:conventions}, in which
\be
\gamma_3 \ov{\xi}_i = \left(\xi^i\right)^\ast.
\ee
Furthermore, we can diagonalize $e_{ij}$ with a global $SU(4)_R$ rotation:
\be
e_{ij} = e^{i \delta}\left(U^T \tilde e ~\! U \right)_{ij} ,
\ee
where $U \in SU(4)_R$, $\delta$ is a real phase, and $\tilde e_{ij}$ is diagonal with real, non-negative entries. The dependence on $\delta$ can be absorbed into the constant value of $\psi$. We choose to write $\tilde e_{ij}$ as
\be
\tilde e_{ij} = \mathrm{diag} \left(b_1, b_2, b_3, b_4 \right), \quad b_i \geq 0.
\ee
So, after rotating $\xi^i$ by $U$, the dilatino constraint is
\be
\left(\tilde\xi^i\right)^\ast =  \tilde e_{ij} \tilde\xi^j.
\ee

It is worth noting that the analysis so far has proceeded similarly to that of~\cite{D'Hoker:2006uv}. Following their notation, let $\beta_{(i)}$, $i = 1, \ldots, 4$, be the eigenvectors of $\tilde e$, satisfying
\be
\tilde e \cdot \beta_{(i)} = b_i \beta_{(i)}.
\ee
In components, $\beta_{(i)}^j = \delta^j{}_i$. Only if one (or more) $b_i = 1$, can we solve the dilatino condition with a constant spinor of the form
\be\label{eq:conservedspinors}
\tilde \xi_{(i)}^j =  \zeta \beta_{(i)}^j,
\ee
where $\zeta$ is any two-component Majorana spinor:
\be
\zeta^\ast = \zeta.
\ee
Such a spinor is the minimal spinor representation of $Spin(1,2)$.

As there can be at most four unit eigenvalues, the maximal amount of supersymmetry under our assumptions conserves $8$ real supercharges, or a three-dimensional $\mathcal{N} = 4$ supersymmetry. We will also find solutions with $\mathcal{N} = 1$ and $\mathcal{N} = 2$ supersymmetry, while $\mathcal{N} = 3$ will be ruled out. 

Lastly, we turn to the auxiliary condition. Supersymmetry requires that the variations of $\chi^{ij}{}_k$ generated by spinors of the form~\C{eq:conservedspinors} vanish. We denote these variations as $\delta_{(l)} \chi^{ij}{}_k$, where $l$ takes only those values for which $b_l = 1$. Then, supersymmetry requires:
\be\label{eq:auxcondition}
\delta_{(l)}\chi^{ij}{}_k \propto q' b_k \epsilon^{ij}{}_{kl} + 2 D^{ij}{}_{kl} + {1 \over 6}|q|^2 \left( 3b_k^2 + 1 - b^2 \right) \delta^{[i}_k\delta^{j]}_l = 0.
\ee 
In this, we have defined:
\be
\epsilon^{ij}{}_{kl} = \epsilon^{ijmn} \delta_{km} \delta_{ln}, \quad b^2 = \sum\limits_{i=1}^4 b_i^2, \quad q =  e^{2i\psi}\left( {i \ov \tau' \over \mathrm{Im} ~\tau} \right).
\ee
And, despite repeated indices, there is no sum over $k$ in~\C{eq:auxcondition}.

Let's for the moment assume that only one of the $b_i = 1$. We choose $b_1 =1$, while $b_{i \neq 1} \neq 1$. This will correspond to preservation of two supercharges or $\mathcal{N} = 1$ supersymmetry in three-dimensions. In this case, the auxiliary condition presents no further constraints; instead, it only fixes the values of $D^{ij}{}_{kl}$:
\be
D^{ij}{}_{k 1} = -{1 \over 2} q'b_k\epsilon^{ij}{}_{k1} -{1 \over 12} |q|^2 \left( 3b_k^2 +1 -b^2 \right) \delta^{[i}_k\delta^{j]}_1.
\ee
Note that this is sufficient to fix all the values of $D^{ij}{}_{kl}$, as there are $20$ independent equations and $20$ independent components to $D^{ij}{}_{kl}$. A particularly simple solution can be found when all $b_{i \neq 1} = 0$. In this case, $D^{ij}{}_{kl} = 0$. This is related to the $\mathcal{N} = 1$ example discussed in~\cite{D'Hoker:2006uv}, differing only in that we have allowed the theta angle to vary as well. As far as we can tell, this background has not appeared before in the literature on Janus configurations.

If, instead, at least two of the $b_i = 1$, which we choose as $b_1 = b_2 = 1$, then we will preserve four or more supercharges, i.e. $\mathcal{N} \geq 2$. This will further constrain the $y$-dependence of $\tau$ due to a consistency condition. Specifically, we have the simultaneous conditions:
\be\label{eq:foursupercharges}
\bes
q' b_k \epsilon^{ij}{}_{k1} + 2 D^{ij}{}_{k1} + {1 \over 6}|q|^2 \left( 3b_k^2 + 1 - b^2 \right) \delta^{[i}_k\delta^{j]}_1 &= 0,\cr
q' b_k \epsilon^{ij}{}_{k2} + 2 D^{ij}{}_{k2} + {1 \over 6}|q|^2 \left( 3b_k^2 + 1 - b^2 \right) \delta^{[i}_k\delta^{j]}_2 &= 0.\cr
\ees
\ee
It is not possible to solve these equations with arbitrary $b_3$, $b_4$, and $q$. To see this, we use the reality condition on $D^{ij}{}_{kl}$
\be
\left( D^{ij}{}_{kl} \right)^\ast = D^{kl}{}_{ij},
\ee
which for a certain choice of $i,j,k$ in~\C{eq:foursupercharges} yields:
\be
\left(q'\right)^\ast b_3 = q' b_4.
\ee
Again, our initial assumptions preclude $q' = 0$ everywhere, so we must have $b_3 = b_4$. This rules out the possibility of $\mathcal{N} = 3$ supersymmetry. Additionally, if $b_3 = b_4 \neq 0$, we must also have $q'$ real. Let's examine this condition, which requires that the imaginary part of $q$ is constant:
\be
\mathrm{Im}~q = 2 \psi' \cos \left( 2\psi \right) -\sin \left( 2\psi \right) {d \over dy} \left( \log g_{YM}^2 \right) = \mathrm{constant}.
\ee
This is the same condition arrived at in~\cite{Gaiotto:2008sd}. As described there, if we seek solutions in which both $\theta'$ and $(g^2)'$ tend to zero as $y \to \pm \infty$, then the constant is actually zero, and we arrive at
\be
\tau(y) = \tau_0 + 4 \pi D e^{-2i \psi}.
\ee
It is worth noting that the configuration described in~\cite{Gaiotto:2008sd} corresponds to all $b_i = 1$.

As we have demonstrated, the known supersymmetric backgrounds of the $\mathcal{N} = 4$ theory fit into the framework of off-shell conformal supergravity. Furthermore, the known backgrounds rely only on a subset of the background fields in the conformal supergravity multiplet, suggesting that a much broader class of backgrounds is possible. A systematic study of the supersymmetry conditions~\C{eq:susyconditions} along the lines of~\cite{Dumitrescu:2012ha,Gupta:2012cy,Klare:2013dka} should reveal these backgrounds, and we hope they may lead to new insights.

\section*{Acknowledgements}
It is my pleasure to thank C. Cordova and S. Sethi for useful discussions. This work was supported, in part, by NSF Grant No.\ PHY-1316960.

\newpage

\appendix
\section{Spinor Conventions}\label{app:conventions}
We work in Lorentzian signature in four dimensions and define $\eta_{ab} = \mathrm{diag}\left(-1, 1,1,1 \right)$. The Clifford algebra is
\be
\acom{\gamma_a}{\gamma_b}_\alpha{}^\beta = 2 \eta_{ab} \delta_\alpha{}^\beta,
\ee
where $\alpha$, $\beta = 1, \ldots 4$ are indices for a Dirac spinor of $Spin(1,3)$. The gamma matrices yield a representation of the $Spin(1,3)$ generators acting on spinors according to
\be
M_{a b} = -{i \over 4} \com{\gamma_a}{\gamma_b} \equiv -{i \over 2} \gamma_{ab}.
\ee
The Dirac representation is reducible into Weyl spinors, eigenstates of the chirality operator
\be
\gamma_{(5)} = i\gamma_0\gamma_1 \gamma_2 \gamma_3,
\ee
which has the properties
\be
\acom{\gamma_{(5)}}{\gamma_a} = 0, \quad \com{\gamma_{(5)}}{M_{ab}} = 0, \quad \gamma_{(5)}^2 = 1, \quad \mathrm{tr}~ \gamma_{(5)} = 0.
\ee
Spinors with eigenvalue $+1$ under $\gamma_{(5)}$ are said to have positive chirality, while those with eigenvalue $-1$ have negative chirality.

In four Lorentzian dimensions, the Weyl spinor is a complex representation whose conjugate is isomorphic to the spinor of opposite chirality. In particular there exist matrices $B$ and $C$ intertwining the equivalent representations of the Clifford algebra:
\be
B^{-1} \gamma_a B = \gamma_a^\ast, \quad C \gamma_a C^{-1}  = -\gamma_a^T.
\ee
These matrices satisfy
\be
B^{-1} \gamma_{(5)} B = -\gamma_{(5)}^\ast, \quad B^\ast B = 1, \quad C \gamma_{(5)} C^{-1}  = \gamma_{(5)}^T, \quad 
C^T = -C.
\ee
The first property demonstrates that the complex conjugate of a Weyl spinors has the opposite chirality, while the properties of $C$ demonstrate that the Weyl spinor is isomorphic, as a vector space, to its dual.

The matrix $C = C^{\alpha \beta}$ yields a canonical isomorphism between a Weyl spinor and its dual. Specifically, it is used to make $Spin(1,3)$ invariants from two Weyl spinors, $\psi_\alpha$, $\eta_\alpha$ of the same chirality:
\be
\psi \eta \equiv \psi^\alpha \eta_\alpha \equiv \psi_\alpha C^{\alpha \beta} \eta_\beta.
\ee
As $C = -C^T$, some care must be taken when using $C^{\alpha \beta}$ or its inverse $C_{\alpha \beta}$ to raise and lower indices. We define
\be
\psi^\alpha = \psi_\beta C^{\beta \alpha}, \quad \psi_\alpha = \psi^\beta C_{\beta \alpha}, \quad C^{\alpha \beta} C_{\beta \gamma} = \delta^\alpha {}_\gamma.
\ee
When $C$ is not specified, we always sum indices from the upper-left to the lower-right.

Other fermion bilinears, transforming under $Spin(1,3)$ can be formed using anti-symmetric products of gamma matrices, defined such that
\be
\gamma^{a_1 \ldots a_k} = \gamma^{[a_1 \ldots a_k]}.
\ee
Again using two fermions of the same chirality, we can form fermion bilinears as
\be
\psi \gamma^{a_1 \ldots a_k} \eta = \psi^\alpha \left( \gamma^{a_1 \ldots a_k} \right)_\alpha{}^\beta \eta_\beta = \psi_\alpha \left( C\gamma^{a_1 \ldots a_k} \right)^{\alpha \beta} \eta_\beta.
\ee
These transform as anti-symmetric tensors of $Spin(1,3)$. The products $C\gamma^{a_1 \ldots a_k}$ have the following symmetry:
\be
\left( C \gamma^{a_1 \ldots a_k} \right)^T = -\left( -1 \right)^{{k(k+1) \over 2}}\left( C\gamma^{a_1 \ldots a_k} \right).
\ee

We may also form fermion bilinears transforming as anti-symmetric tensors of $Spin(1,3)$ from a fermion $\psi$ and its complex conjugate $\psi^\ast$. Before writing these, we will define
\be
\ov \psi = B \psi^\ast.
\ee
With this definition, we define the bilinears as
\be
 \ov\psi \gamma^{a_1 \ldots a_k} \psi = \ov \psi_\alpha \left( C \gamma^{a_1 \ldots a_k} \right)^{\alpha \beta} \psi_\beta = \psi^\ast_\alpha \left( B^T C \gamma^{a_1 \ldots a_k} \right)^{\alpha \beta} \psi_\beta.
\ee
The matrix $B^T C$ is often called $A$, which we can choose to be hermitian~\cite{Sohnius:1985qm}. Then, the products $B^TC\gamma^{a_1 \ldots a_k}$ have the following hermiticity:
\be
\left(B^TC\gamma^{a_1 \ldots a_k}\right)^\dagger = \left(-1\right)^{{k(k+1)\over 2}} B^TC\gamma^{a_1 \ldots a_k}.
\ee

At some point, it will be convenient for us to work with a specific basis in which the gamma matrices take the form:
\be
\gamma_0 = i \sigma^2 \otimes \sigma^2, \quad \gamma_1 = \sigma^1 \otimes \sigma^2, \quad \gamma_2 = \sigma^3 \otimes \sigma^2, \quad \gamma_3 = 1 \otimes \sigma^1,
\ee
where $\sigma^i$ are the Pauli matrices. This is a Weyl basis, in which $\gamma_{(5)}$ is diagonal:
\be
\gamma_{(5)} = 1 \otimes \sigma^3.
\ee
And the intertwiners are given by
\be
B = \gamma_3, \quad C = i \gamma_3 \gamma_0, \quad B^TC = i\gamma_0.
\ee

Our spinors will also transform under an internal $SU(4)_R$ symmetry. We denote the $\mathbf{4}$ representation of $SU(4)_R$ with raised indices $i, j, \ldots = 1, \ldots, 4$. The conjugate representation $\mathbf{\ov 4}$ is denoted with lowered indices. In addition to the invariant tensors $\delta^i{}_j$ and $\delta_i{}^j$, $SU(4)$ also possess the totally antisymmetric invariant tensors $\epsilon_{ijkl}$ and $\epsilon^{ijkl}$. These tensors can be used to raise and lower $\mathbf{4}$ and $\ov{\mathbf{4}}$ indices. A relevant example pertains to the $\mathbf{6}$ representation, which is given by an antisymmetric, two-component tensor $\phi^{ij} = -\phi^{ji}$. We lower these indices with $\epsilon_{ijkl}$ according to the convention
\be
\phi_{ij} = {1 \over 2} \epsilon_{ijkl} \phi^{kl}.
\ee

\section{Gauge Fixing}\label{app:gaugefixing}
In~\cite{Bergshoeff:1980is}, $\Phi_{\ul \alpha} = \left( \Phi_1, \Phi_2 \right)$ satisfies the $SU(1,1)$ invariant constraint:
\be
|\Phi_1|^2 - |\Phi_2|^2 = 1.
\ee
By changing basis, we can cast this in the language of $SL(2, \mathbb{R})$:
\be
\wt \Phi_1 = {1 \over \sqrt{2}} \left( \Phi_1 + \Phi_2 \right), \quad \wt \Phi_2 = {i\over \sqrt{2}} \left( \Phi_1 - \Phi_2\right).
\ee
In terms of these variables, the constraint reads
\be
i \wt\Phi_1 \wt\Phi_2^\ast -i \wt\Phi^\ast_1 \wt\Phi_2 = 1.
\ee
Then, the matrix
\be
U = \sqrt{2} \begin{pmatrix} \mathrm{Re}~\wt\Phi_1 & \mathrm{Im}~\wt\Phi_1 \\ \mathrm{Re}~\wt\Phi_2 & \mathrm{Im}~\wt\Phi_2 \end{pmatrix}
\ee
is an element of $SL(2, \mathbb{R})$. Transformations of $\wt \Phi_{\ul \alpha}$ under $SL(2,\mathbb{R}) \times U(1)$  act by left multiplication with an $SL(2,\mathbb{R})$ matrix and right multiplication with a local $U(1)$ matrix
\be
U \to \begin{pmatrix} a & b \\ c & d \end{pmatrix} U \begin{pmatrix} \cos \theta(x) & -\sin \theta(x) \\ \sin\theta(x) & \cos\theta(x) \end{pmatrix}, \quad a,b,c,d \in \mathbb{R}, \quad ad - bc = 1.
\ee
The coupling parameter $\tau$ is related to $\wt\Phi_{\ul \alpha}$ by
\be
\tau(\wt\Phi) = {\wt \Phi_1^\ast \over \wt\Phi_2^\ast}.
\ee
With this identification, $\tau$ is invariant under $U(1)$ and has the standard $SL(2, \mathbb{R})$ transformation:
\be
\tau \to \left({a \tau + b \over c \tau +d}\right).
\ee

A convenient gauge-fixing is to choose $\mathrm{Re}~\wt\Phi_2 = 0$. This gauge choice is invariant under the subgroup of $SL(2,\mathbb{R}) \times U(1)$ transformations in which the $U(1)$ parameter $\theta(x)$ is given by
\be
e^{i \theta(x)} = {|c\tau + d| \over c\tau + d}.
\ee
This is the same transformation found in two other contexts in~\cite{Kapustin:2006pk,Harvey:2007ab}.

After performing this gauge fixing, some of the formulas in~\cite{Bergshoeff:1980is,deRoo:1984zyh,deRoo:1985np} are simplified. While we have not presented the original formulas, we list some of the conversions below for the reader interested in comparing to previous literature.
\be
\bes
\Phi = \Phi^\ast = \Phi_1^\ast - \Phi_2^\ast  &= {1 \over \sqrt{ \mathrm{Im} \tau}}, \cr
D_\mu \Phi^{\ul\alpha} D^\mu \Phi_{\ul \alpha} &= - {\partial_\mu \tau \partial^\mu \ov \tau \over 4 \left( \mathrm{Im} \tau \right)^2}.
\ees
\ee
The $\tau$-dependent terms in the supersymmetry variations of the background fermions are:
\be
\bes
a_\mu &= -{ \partial_\mu \left( \tau + \ov \tau \right) \over 4 \mathrm{Im} \tau}, \cr
\epsilon^{\ul \alpha \beta} \Phi_{\ul \alpha} D_\mu \Phi_{\ul \beta} &= -i a_\mu -{i \over 2} \partial_\mu \log \left(\mathrm{Im}~\! \tau\right).
\ees
\ee

\newpage
\bibliographystyle{utphys}
\bibliography{bib}
\end{document}